\documentclass[a4paper,12pt]{article}
\usepackage{graphicx}
\usepackage{float}
\usepackage{amsmath}
\usepackage{amssymb}
\usepackage{amsfonts}
\usepackage{theorem}
\usepackage{amscd}
\usepackage{latexsym}

\usepackage{eurosym}

\numberwithin{equation}{section}

\newtheorem{theo}{Theorem}[section]
\newtheorem{lem}{Lemma}[section]

\newtheorem{defi}{Definition}[section]

\renewcommand{\thefootnote}{\fnsymbol{footnote}}

\newcommand{\prob}{\mathbb{P}}
\newcommand{\esp}{\mathbb{E}}

\newcommand{\R}{\mathbb R}

\newcommand{\rhotau}{\rho_\tau}

\DeclareMathOperator*{\argmin}{argmin}

\newif\ifnotes
\notestrue
\newcounter{mnotecount}[section]

\begin{document}

\begin{center}

{\sc \Large Sequential Quantile Prediction of Time Series \\
\vspace{0.7cm}}

G\'erard BIAU $^{\mbox{\footnotesize a},}\footnote{Partially supported by the French ``Agence Nationale pour la Recherche''
under grant ANR-09-BLAN-0051-02 ``CLARA''. Research carried out within the INRIA project ``CLASSIC'' hosted by Ecole Normale Sup{\'e}rieure and CNRS.}$ and Beno\^{i}t PATRA $^{\mbox{\footnotesize a,b,}}\footnote{Corresponding
author.}$
\vspace{0.5cm}

$^{\mbox{\footnotesize a}}$ LSTA\\Universit\'e Pierre et Marie Curie -- Paris VI\\
Bo\^{\i}te 158,   175 rue du Chevaleret\\
75013 Paris, France\\
\smallskip
\textsf{gerard.biau@upmc.fr}\\
\bigskip
$^{\mbox{\footnotesize b}}$
LOKAD SAS\\
70 rue Lemercier\\
75017 Paris, France\\
\smallskip
\textsf{benoit.patra@lokad.com}\\

\vspace{0.5cm}

\end{center}

\begin{abstract}
\noindent {\rm Motivated by a broad range of potential applications, we address the quantile prediction problem of real-valued time series. We present a sequential quantile forecasting model based on the combination of a set of elementary nearest neighbor-type predictors called ``experts'' and show its consistency under a minimum of conditions. Our approach builds on the methodology developed in recent years for prediction of individual sequences and exploits the quantile structure as a minimizer of the so-called pinball loss function. We perform an in-depth analysis of real-world data sets
and show that this nonparametric strategy generally outperforms standard quantile prediction methods.\\

\medskip

\medskip

}

\end{abstract}

\renewcommand{\thefootnote}{\arabic{footnote}}

\setcounter{footnote}{0}
\section{Introduction}
Forecasting the future values of an observed time series is an important problem, which has been an area of considerable activity in recent years. The application scope is vast, as time series prediction applies to many fields, including problems in genetics, medical
diagnoses, air pollution forecasting, machine condition
monitoring, financial investments, production planning, sales forecasting and stock controls. \\

To fix the mathematical context, suppose that at each time instant $n=1, 2, \hdots$,
the forecaster (also called the predictor hereafter) is
asked to guess the next outcome $y_n$ of a sequence of real numbers
$y_1, y_2, \hdots$ with knowledge of the past $y_1^{n-1}=(y_1,
\hdots,y_{n-1})$ (where $y_1^0$ denotes the empty string). Formally, the
strategy of the predictor is a sequence $g=\{g_n\}_{n=1}^{\infty}$ of
forecasting functions
\[
   g_n: \mathbb R^{n-1} \to \mathbb R
\]
and the prediction formed at time $n$ is just  $g_n(y_1^{n-1})$. Throughout the paper we will suppose that
$y_1$, $y_2,\hdots$  are realizations of random variables $Y_1$, $Y_2, \hdots$ such that the stochastic process $\{Y_n\}_{-\infty}^{\infty}$ is jointly stationary and ergodic.\\

Many of the statistical techniques used in time series prediction are those of regression analysis, such as classical least square theory, or are adaptations or analogues of them. These forecasting schemes are typically concerned with finding a function $g_n$ such that the prediction $g_n(Y_1^{n-1})$ corresponds to the conditional mean of $Y_n$ given the past sequence $Y_{1}^{n-1}$, or closely related quantities. Many methods have been developed for this purpose, ranging from parametric approaches such as AR($p$) and ARMA($p$,$q$) processes (Brockwell and Davies
\cite{BRO1}) to more involved nonparametric methods (see for example
Gy\"orfi et al. \cite{GYO1} and Bosq \cite{BOS1}
for a review and references).\\

On the other hand, while these estimates of the conditional mean serve their purpose, there exists a large area of problems where the forecaster is more interested in estimating conditional quantiles and prediction intervals, in order to know other features of the conditional distribution. There is now a fast pace growing literature on quantile regression (see Gannoun, Saracco and Yu \cite{GAN1} for an overview and references) and considerable practical experience with forecasting methods based on this theory. Economics makes a persuasive case for the value of going beyond models for the conditional mean (Koenker and Allock \cite{KOE3}). In financial mathematics and financial risk management, quantile regression is intimately linked to the $\tau$-Value at Risk (VaR), which is defined as the $(1-\tau)$-quantile of the portfolio. For example, if a portfolio of stocks has a one-day 5\%-VaR of \euro 1 million, there is a 5\% probability that the portfolio will fall in value by more than \euro 1 million over a one day period (Duffie and Pan \cite{DAR1}). More generally, quantile regression methods have been deployed in social sciences, ecology, medicine and manufacturing process management. For a description, practical guide and extensive list of references on these methods and related methodologies, we refer the reader to the monograph of Koenker \cite{KOE1}.\\

Motivated by this broad range of potential applications, we address in this paper the quantile prediction problem of real-valued time series. Our approach is nonparametric in spirit and breaks with at least three aspects of more traditional procedures. First, we do not require the series to necessarily satisfy the classical statistical assumptions for bounded, autoregressive or Markovian processes. Indeed, our goal is to show powerful
consistency results under a strict minimum of conditions. Secondly, building on the methodology developed in recent years for prediction of individual sequences, we present a sequential quantile forecasting model based on the combination of a set of elementary nearest neighbor-type predictors called ``experts''. The paradigm of prediction with expert advice was first introduced in the theory of machine learning as a model of online learning in the 1980-early 1990s, and it has been extensively investigated ever since (see the monograph of Cesa-Bianchi and Lugosi \cite{CES1} for a comprehensive introduction to the domain). Finally, in opposition to standard nonparametric approaches, we attack the problem by  fully exploiting the quantile structure as a minimizer of the so-called pinball loss function (Koenker and Basset \cite{KOE2}).\\

The document is organized as follows. After some basic recalls in Section 2, we present in Section 3 our expert-based quantile prediction procedure and state its consistency under a minimum of conditions. We perform an in-depth analysis of real-world data sets
and show that the nonparametric strategy we propose is faster and generally outperforms traditional methods in terms of average prediction errors (Section 4). Proofs of the results are postponed to Section 5.

\section{Consistent quantile prediction}
\subsection{Notation and basic definitions}
\begin{figure}[!h]
\label{plf}
\begin{center}
\includegraphics[scale = 0.5]{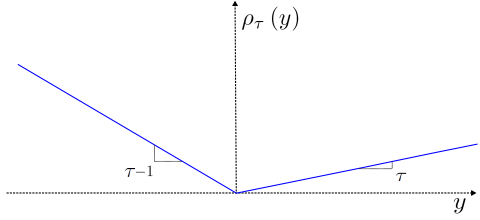}
\end{center}
\caption{Pinball loss function $\rhotau$.}
\end{figure}

Let $Y$ be a real-valued random variable with distribution function $F_Y$, and let $\tau \in (0,1)$. Recall that the generalized inverse of $F_Y$
$$F^{\leftarrow}_Y(\tau)=\inf \{t \in \mathbb R : F_Y(t) \geq \tau\}$$
is called the quantile function of $F_Y$ and that the real number $q_{\tau}=F^{\leftarrow}_Y(\tau)$ defines the $\tau$-quantile of $F_Y$ (or $Y$).
The basic strategy behind quantile estimation arises from the observation that minimizing the $\ell_1$-loss function yields the median. Koenker and Basset \cite{KOE2} generalized this idea and characterized the $\tau$-quantile by tilting the absolute value function in a suitable fashion.
\begin{lem}
\label{tilt}
Let $Y$ be an integrable real-valued random variable and, for $\tau \in (0,1)$, let the map
$$\rho_{\tau}(y)= y(\tau - \mathbf 1_{[y \leq 0]}).$$
Then the quantile $q_{\tau}$ satisfies the property
\begin{equation}
\label{wep}
q_{\tau} \in \argmin_{q \in \mathbb R} \mathbb{E}\left[\rhotau(Y-q) \right].
\end{equation}
Moreover, if $F_Y$ is (strictly) increasing, then the minimum is unique, that is
$$
\{q_{\tau}\} = \argmin_{q \in \mathbb R} \mathbb{E}\left[\rhotau(Y-q) \right].
$$
\end{lem}
We have not been able to find a complete proof of this result, and we briefly state it in Section 5. The function $\rho_{\tau}$, shown in Figure \ref{plf}, is called the pinball function. For example, for $\tau=1/2$, it yields back the absolute value function and, in this case, Lemma \ref{tilt} just expresses the fact that the median $q_{1/2}=F^{\leftarrow}(1/2)$ is a solution of the minimization problem
$$q_{1/2} \in \argmin_{q \in \mathbb R} \mathbb{E}|Y-q|.$$

These definitions may be readily extended to pairs $(X,Y)$ of random variables with conditional distribution $F_{Y|X}$. In this case, the conditional quantile $q_{\tau}(X)$ is the measurable function of $X$ almost surely (a.s.) defined by
$$q_{\tau}(X)=F_{Y|X}^{\leftarrow}(\tau)=\inf \{t \in \mathbb R : F_{Y|X}(t) \geq \tau\},$$
and, as in Lemma \ref{tilt}, it can be shown that for an integrable $Y$
\begin{equation}
\label{wepp}
q_{\tau}(X) \in \argmin_{q(.)} \mathbb{E}_{\mathbb P_{Y|X}}\left[\rhotau\left(Y-q(X)\right) \right],
\end{equation}
where the infimum is taken over the set of all measurable real-valued functions and the notation $\mathbb{E}_{\mathbb P_{Y|X}}$ stands for the conditional expectation of $Y$ with respect to $X$. We note again that if $F_{Y|X}$ is a.s. increasing, then the solution of (\ref{wepp}) is unique and equals $q_{\tau}(X)$ a.s.  In the sequel, we will denote by $\mathcal Q_{\tau}(\mathbb P_{Y|X})$ the set of solutions of the minimization problem (\ref{wepp}), so that $q_{\tau}(X) \in \mathcal Q_{\tau}(\mathbb P_{Y|X})$ and $\{q_{\tau}(X)\}=\mathcal Q_{\tau}(\mathbb P_{Y|X})$ when the minimum is unique.

\subsection{Quantile prediction}
In our sequential version of the quantile prediction problem, the forecaster observes one after another the realizations $y_1, y_2, \hdots$ of a stationary and ergodic random process $Y_1, Y_2, \hdots$ At each time $n=1, 2, \hdots$, before the $n$-th value of the sequence is revealed, his mission is to guess the value of the conditional quantile
$$q_{\tau}(Y_1^{n-1})=F_{Y_n|Y_1^{n-1}}^{\leftarrow}(\tau)=\inf \{t \in \mathbb R : F_{Y_n|Y_1^{n-1}}(t) \geq \tau\},$$
on the basis of the previous $n-1$ observations $Y_1^{n-1}=(Y_1,\hdots,Y_{n-1})$ only. Thus, formally, the
strategy of the predictor is a sequence $g=\{g_n\}_{n=1}^{\infty}$ of quantile prediction functions
\[
   g_n: \mathbb R^{n-1} \to \mathbb{R}
\]
and the prediction formed at time $n$ is just  $g_n(y_1^{n-1})$. After $n$ time instants, the (normalized) cumulative quantile loss on the string $Y_1^n$ is
$$L_n(g)=\frac{1}{n}\sum_{t=1}^{n}{\rho_{\tau}\left( Y_t - g_t (Y_{1}^{t-1}) \right) }.$$
Ideally, the goal is to make $L_n(g)$ small. There is, however, a fundamental limit for the quantile predictability, which is determined by a result of Algoet \cite{ALG2}: for any quantile prediction strategy $g$ and jointly stationary ergodic process $\{Y_n\}_{-\infty}^{\infty}$,
\begin{equation}
\label{prop:alg}
\liminf_{n \to \infty} L_n(g) \geq L^{\star} \quad \mbox{a.s.},
\end{equation}
where
$$L^{\star}=\mathbb{E}\left[\min_{q(.)}\mathbb E_{\prob_{Y_0 | Y_{-\infty}^{-1}}} \left[\rhotau \left(Y_0 - q(Y_{-\infty}^{-1} )\right)\right]\right]$$
is the expected minimal quantile loss over all quantile estimations of $Y_0$ based on the infinite past observation sequence $Y_{-\infty}^{-1}=(\hdots, Y_{-2}, Y_{-1})$. Generally, we cannot hope to design a strategy whose prediction error exactly achieves the lower bound $L^{\star}$. Rather, we require that $L_n(g)$ gets arbitrarily close to $L^{\star}$ as $n$ grows. This gives sense to the following definition:
\begin{defi}
A quantile prediction strategy $g$ is called consistent with respect to a class $\mathcal C$ of stationary and ergodic processes $\{Y_n\}_{-\infty}^{\infty}$ if, for each process in the class,
$$ \lim_{n \to \infty} L_n(g)=L^{\star}\quad \mbox{a.s.}$$
\end{defi}
Thus, consistent strategies asymptotically achieve the
best possible loss for all processes in the class. In the context of prediction with squared loss, Gy\"orfi and Lugosi  \cite{GYO3}, Nobel \cite{NOB1}, Gy\"orfi and Ottucs\'ak \cite{GYO5} and Biau et al. \cite{BIA1} study various sequential prediction strategies, and state their consistency under a minimum of assumptions on the collection $\mathcal C$ of stationary and ergodic processes. Roughly speaking, these methods consider several ``simple'' nonparametric estimates (called experts in this context) and combine them at time $n$ according to their past performance. For this, a probability distribution on the set of experts is generated,
where a ``good'' expert has relatively large weight, and the average of all experts' predictions is taken with respect to this
distribution. Interestingly, related schemes have been proposed in the context of sequential investment strategies for financial markets. Sequential investment strategies are allowed to use information about the market collected from the past and determine at the beginning of a training period a portfolio, that is, a way to distribute the current capital among the available assets. Here, the goal of the investor is to maximize his wealth in the long run, without knowing the underlying distribution generating the stock prices. For more information on this subject, we refer the reader to Algoet \cite{ALG1}, Gy\"orfi and Sch\"afer \cite{GYO6}, Gy\"orfi, Lugosi and Udina \cite{GYO4}, and Gy\"orfi, Udina and Walk \cite{GYO8}.
\\

Our purpose in this paper will be to investigate an expert-oriented strategy for quantile forecasting. With this aim in mind, we define in the next section a quantile prediction strategy, called nearest neighbor-based strategy, and state its consistency with respect to a large class of stationary and ergodic processes.
\section{A nearest neighbor-based strategy}
The quantile prediction strategy is defined at each time instant as a convex combination of elementary predictors (the so-called experts), where the
wei\-gh\-ting coefficients depend on the past performance of each
elementary predictor. To be more precise, we first define an infinite array of experts
$h_n^{(k,\ell)}$, where $k$ and $\ell$ are positive integers. The integer $k$ is the length of the past observation vectors
being scanned by the elementary expert and, for each
$\ell$, choose $p_{\ell} \in (0,1)$
such that
\begin{equation*}
\lim_{\ell \to \infty} p_{\ell}=0 \label{eq:pl}\,,
\end{equation*}
and set
$$\bar \ell=\lfloor p_{\ell} n\rfloor$$
(where $\lfloor . \rfloor$ is the floor function).
At time $n$, for fixed $k$ and $\ell$ ($n > k + \bar \ell +1)$, the
expert searches for the $\bar \ell$ nearest neighbors (NN) of the
last seen observation $y_{n-k}^{n-1}$ in the past and
predicts the quantile accordingly. More precisely, let
\begin{eqnarray*}
J_{n}^{(k,\ell)}\!\!
=\!\!\!\!&\big\{\!\!\!\!& k<t<n  : y_{t-k}^{t-1} \mbox{ is among
the } \bar \ell \mbox{-NN of } y_{n-k}^{n-1} \mbox{ in } y_1^k, \hdots, y_{n-k-1}^{n-2}\big\},
\end{eqnarray*}
and define the elementary predictor $\bar h_n^{(k,\ell)}$ by

\begin{equation*}
\bar h_n^{(k,\ell)}\in \argmin_{q \in \mathbb R} \sum_{t \in J_{n}^{(k,\ell)}}\rhotau{(y_t - q )}
\end{equation*}
if $n > k + \bar \ell +1$, and $0$ otherwise. Next, let the truncation function
\[
 T_a(z) = \left \{ \begin{array}{ll}
       a & \mbox{if $z > a$;} \\
       z & \mbox{if $|z| \le a$;}\\
       -a & \mbox{if $z < -a$,}
                \end{array}   \right.
\]
and let
 \begin{equation}
\label{elexpert}
 h_n^{(k,\ell)} = T_{\min(n^\delta,\ell)}\circ\bar{h}_n^{(k,\ell)},
 \end{equation}
 where $\delta$ is a positive parameter to be fixed later on. We note that the expert  $h_n^{(k,\ell)}$  can be interpreted as a (truncated) $\bar \ell$-nearest neighbor regression function estimate drawn in $\mathbb R^k$ (Gy\"orfi et al. \cite{GYO2}). The proposed quantile prediction algorithm proceeds with an exponential
weighting average of the experts. More formally, let $\{b_{k,\ell}\}$ be a
probability distribution on the set of all pairs $(k,\ell)$ of
positive integers such that for all $k$ and $\ell$, $b_{k,\ell}>0$. Fix a learning parameter
$\eta_n>0$, and define the weights
\[
w_{k,\ell,n}=b_{k,\ell}e^{-\eta_n(n-1)L_{n-1}(h_n^{(k,\ell)})}
\]
and their normalized values
\[
p_{k,\ell,n}=\frac{w_{k,\ell,n}}{\sum_{i,j=1}^{\infty}w_{i,j,n}}.
\]
The quantile prediction strategy $g$ at time $n$ is defined by
\begin{equation}
\label{QPS}
  g_n(y_1^{n-1})
  =
  \sum_{k,\ell=1}^{\infty}p_{k,\ell,n}h_n^{(k,\ell)}(y_1^{n-1}),
\qquad n=1,2,\ldots
\end{equation}
The idea of combining a collection of concurrent estimates was
originally developed in a non-stochastic context for online
sequential prediction from deterministic sequences (Cesa-Bianchi
and Lugosi \cite{CES1}). Following the terminology of
the prediction literature, the combination of different procedures
is sometimes termed aggregation in the stochastic context. The
overall goal is always the same: use aggregation to improve
prediction. For a recent review and an updated list of references,
see Bunea and Nobel
\cite{BUN1}.\\

In order to state consistency of the method, we shall impose the following set of assumptions:

\begin{enumerate}
\item[$(H1)$] One has $\mathbb E[Y_0^2] < \infty$.

\item[$(H2)$] For any vector $\mathbf s \in \mathbb R^k$, the random variable
$\|Y_1^{k}-\mathbf s\|$ has a continuous distribution function.

\item[$(H3)$] The conditional distribution function $F_{Y_0|Y_{-\infty}^{-1}}$ is a.s. increasing.
\end{enumerate}
Condition $(H2)$ expresses the fact that ties occur with probability zero. A discussion on how to deal with ties that may
appear in some cases can be found in \cite{GYO7}, in the related context of portfolio selection strategies. Condition $(H3)$ is mainly technical and ensures that the minimization problem (\ref{wepp}) has a unique solution or, put differently, that the set $\mathcal Q_{\tau}(\mathbb P_{Y_0|Y_{-\infty}^{-1}})$ reduces to the singleton $\{F^{\leftarrow}_{Y_0|Y_{-\infty}^{-1}}(\tau)\}$.\\

We are now in a position to state the main result of the paper.

\begin{theo}
\label{thm:main}
Let $\mathcal C$ be the class of all jointly stationary ergodic
processes $\{Y_n\}_{-\infty}^{\infty}$ satisfying conditions $(H1)$-$(H3)$. Suppose in addition that $n\eta_n \to \infty$ and $n^{2\delta}\eta_n \to 0$ as $n \to \infty$. Then the nearest neighbor
quantile prediction strategy defined above is consistent with respect to $\mathcal{C}$.
\end{theo}
The truncation index $T$ in definition (\ref{elexpert}) of the elementary expert $h_n^{(k,\ell)}$ is merely a technical choice that avoids having to assume that $|Y_0|$ is a.s. bounded. On the practical side, it has little influence on results for relatively short time series. On the other hand, the choice of the learning parameter $\eta_n$ as $1/\sqrt n$ ensures consistency of the method for $0<\delta < \frac{1}{4}$.
\section{Experimental results}
\subsection{Algorithmic settings}
In this section, we evaluate the behavior of the nearest neighbor quantile prediction strategy on real-world data sets and compare its performances to those of standard families of methods on the same data sets.\\

Before testing the different procedures, some precisions on the computational aspects of the presented method are in order. We first note that infinite sums make formula (\ref{QPS}) impracticable. Thus, for practical reasons, we chose a finite grid $(k,\ell) \in \mathcal K \times \mathcal L$ of experts (positive integers), let
\begin{equation}
\label{CQPS}
  g_n(y_1^{n-1})
  =
  \sum_{k \in \mathcal K, \ell \in \mathcal L}p_{k,\ell,n}h_n^{(k,\ell)}(y_1^{n-1}),
\qquad n=1,2,\ldots
\end{equation}
and fixed the probability distribution $\{q_{k,\ell}\}$
as the uniform distribution over the $|\mathcal K| \times |\mathcal L|$ experts. Observing that $h_n^{(k, \ell_1)} = h_n^{(k, \ell_2)}$ and $b_{k,\ell_1} = b_{k,\ell_2}$ whenever $\bar{\ell_1} = \bar{\ell_2}$, formula ($\ref{CQPS}$) may be more conveniently rewritten as
$$g_n(y_1^{n-1}) = \sum_{k \in \mathcal K, \bar{\ell} \in \bar{\mathcal L}}p_{k,\bar{\ell},n} h_n^{(k, \bar{\ell} )}(y_1^{n-1}),$$
where $\bar {\mathcal L}=\{\bar {\ell} : \ell \in \mathcal L\}$. In all subsequent numerical experiments, we chose $\mathcal K=\{1,2,3,\ldots,14\}$ and $\bar{\mathcal L} = \{1,2,3,\ldots,25\}$. \\

Next, as indicated by the theoretical results, we fixed $\eta_n = \sqrt{{1}/{n}}$. For a thorough discussion on the best practical choice of $\eta_n$, we refer to \cite{BIA1}. To avoid numerical instability problems while computing the $p_{k, \bar{\ell} ,n}$, we applied if necessary a simple linear transformation on all $L_n(h_n^{(k,\bar{\ell})})$, just to force these quantities to belong to an interval where the effective computation of $x \mapsto \exp(-x)$ is numerically stable.  \\

Finally, in order to deal with the computation of the elementary experts $(\ref{elexpert})$, we denote by $\lceil . \rceil$ the ceiling function and observe that if $m\times \tau$ is not an integer, then the solution of the minimization problem $\argmin_{b \in \R}\sum_{i=1}^m \rho_\tau\left(y_i -b \right)$ is unique and equals the $\lceil m \times\tau \rceil$-th element in the sorted sample list. On the other hand, if $m\times \tau$ is an integer then the minimum is not unique, but the $m\times \tau$-th element in the sorted sequence may be chosen as a minimizer. Thus, practically speaking, each elementary expert is computed by sorting the sample. The complexity of this operation is $\mbox{O}(\bar{\ell}\log(\bar{\ell}))$---it is almost linear and feasible even for large values of $\bar{\ell}$ . For a more involved discussion, we refer the reader to Koenker \cite{KOE1}. \\

All algorithms have been implemented using the oriented object language C\# 3.0 and .NET Framework 3.5.
\subsection{Data sets and results}

We investigated 21 real-world time series representing daily call volumes entering call centers. Optimizing the staff level is one of the most difficult and important tasks for a call center manager. Indeed, if the staff is overdimensioned, then most of the employees will be inactive. On the other hand, underestimating the staff may lead to long waiting phone queues of customers. Thus, in order to know the right staff level, the manager needs to forecast the call volume series and, to get a more accurate staff level planning, he has to forecast the quantiles of the series.\\

\begin{figure}[!h]
\label{series}
$\begin{array}{cc}
\includegraphics[scale = 0.17]{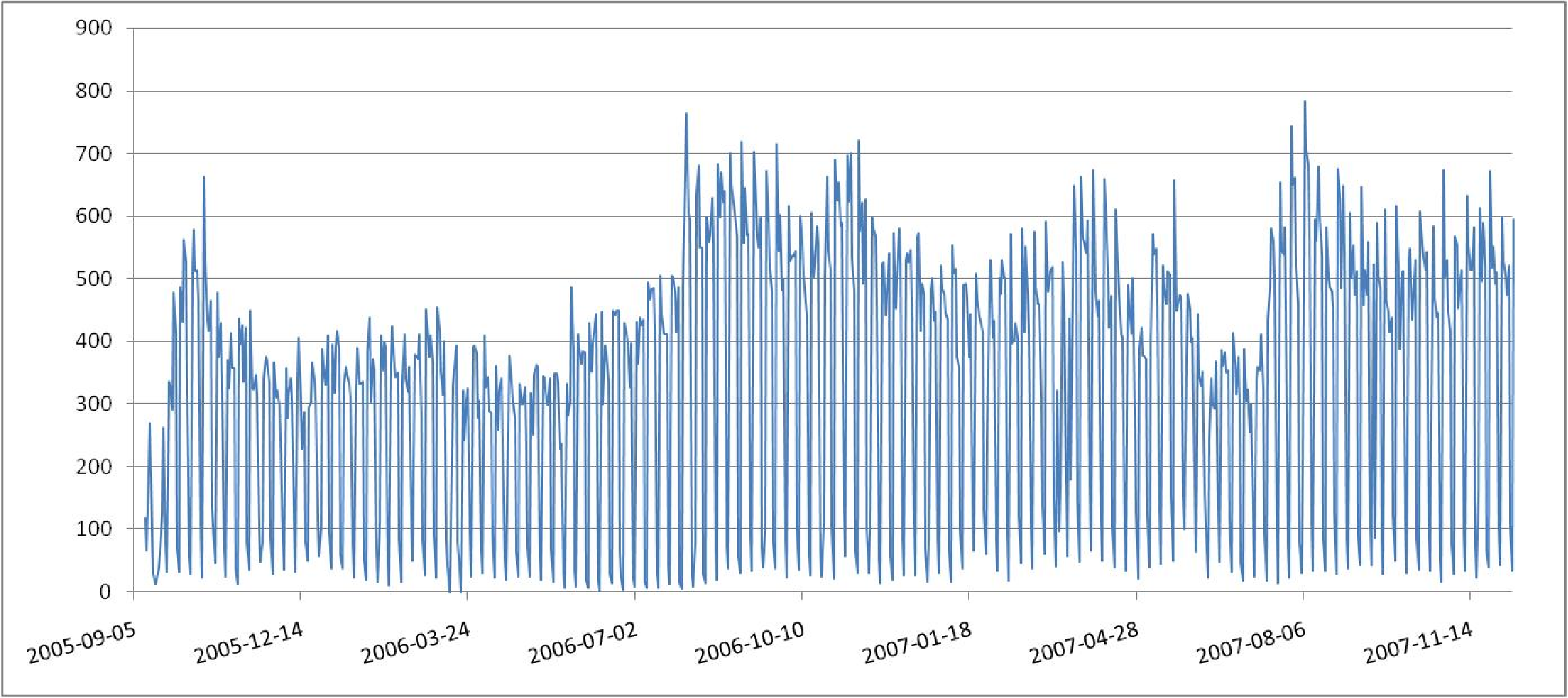} & \includegraphics[scale = 0.17]{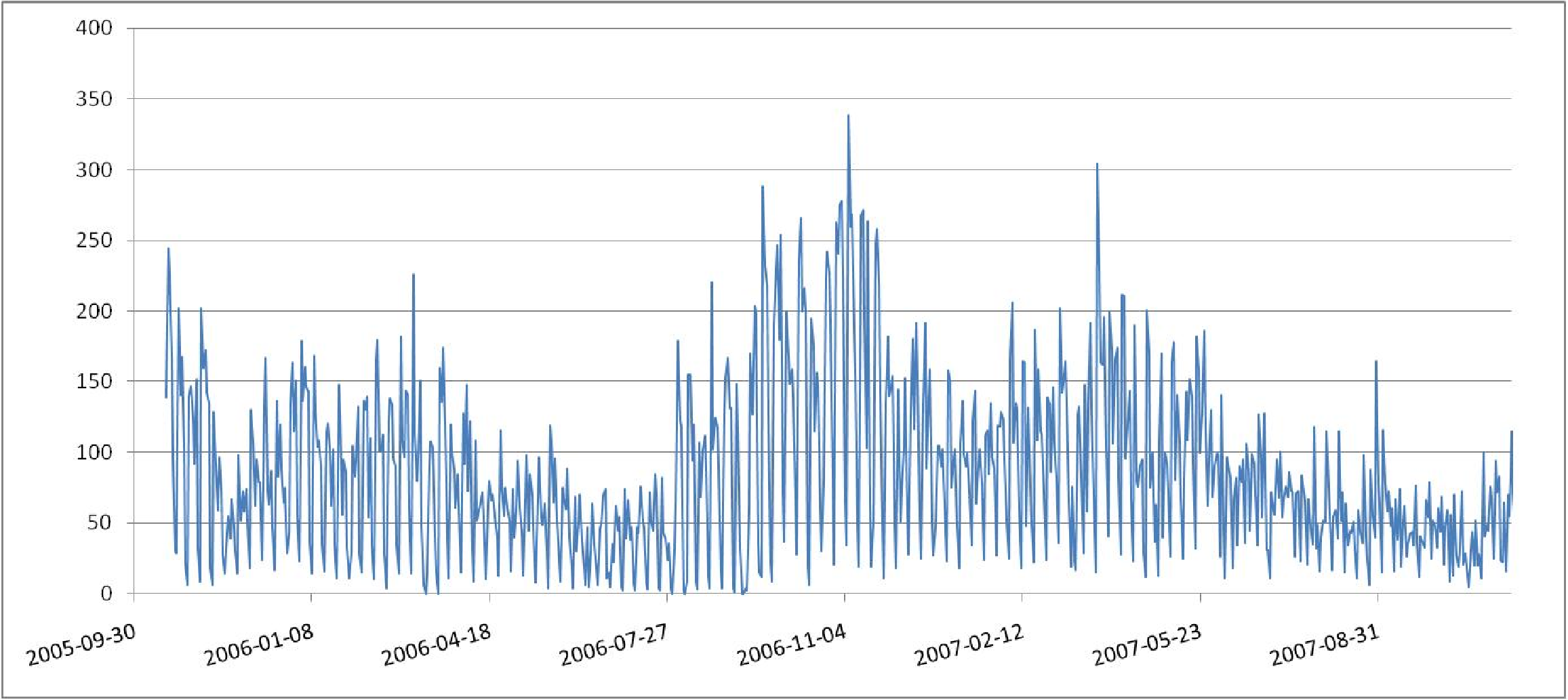} \\
\includegraphics[scale = 0.17]{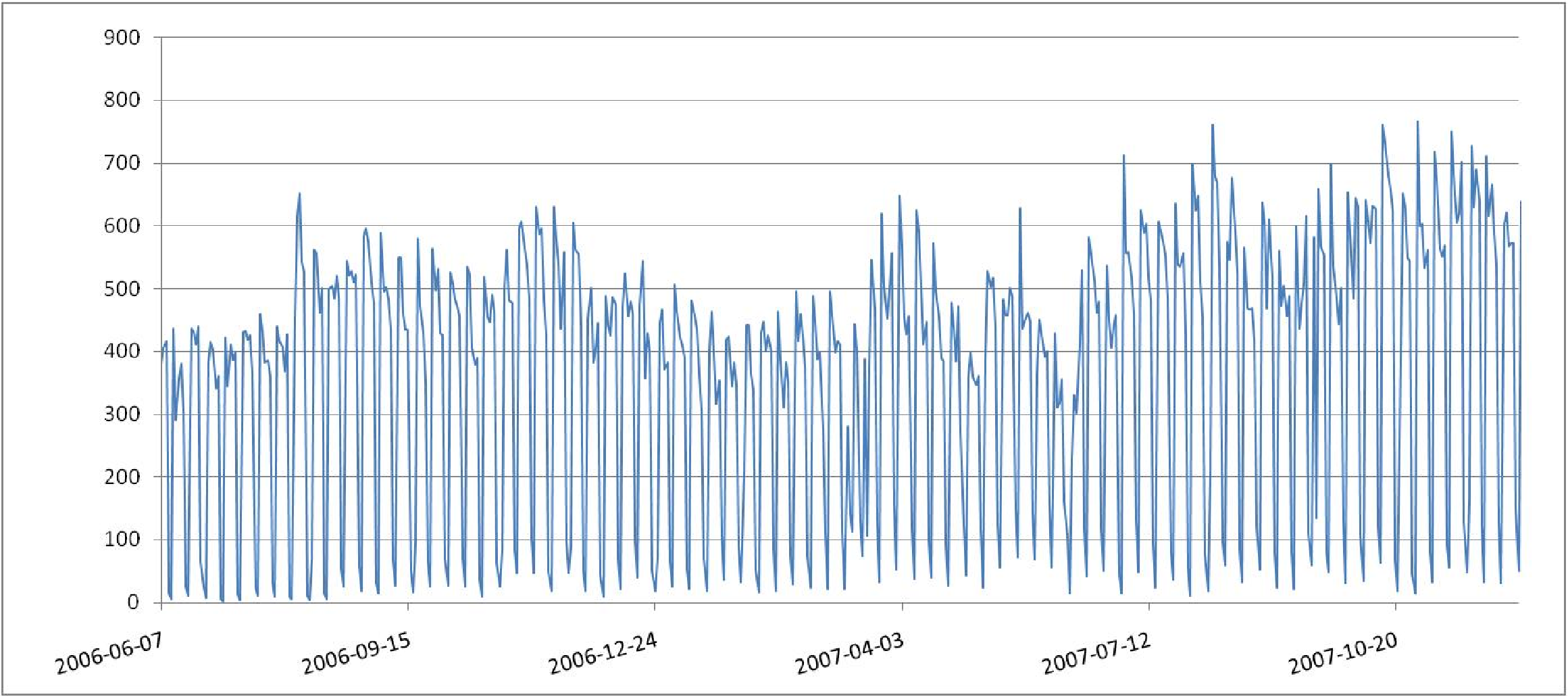} & \includegraphics[scale = 0.17]{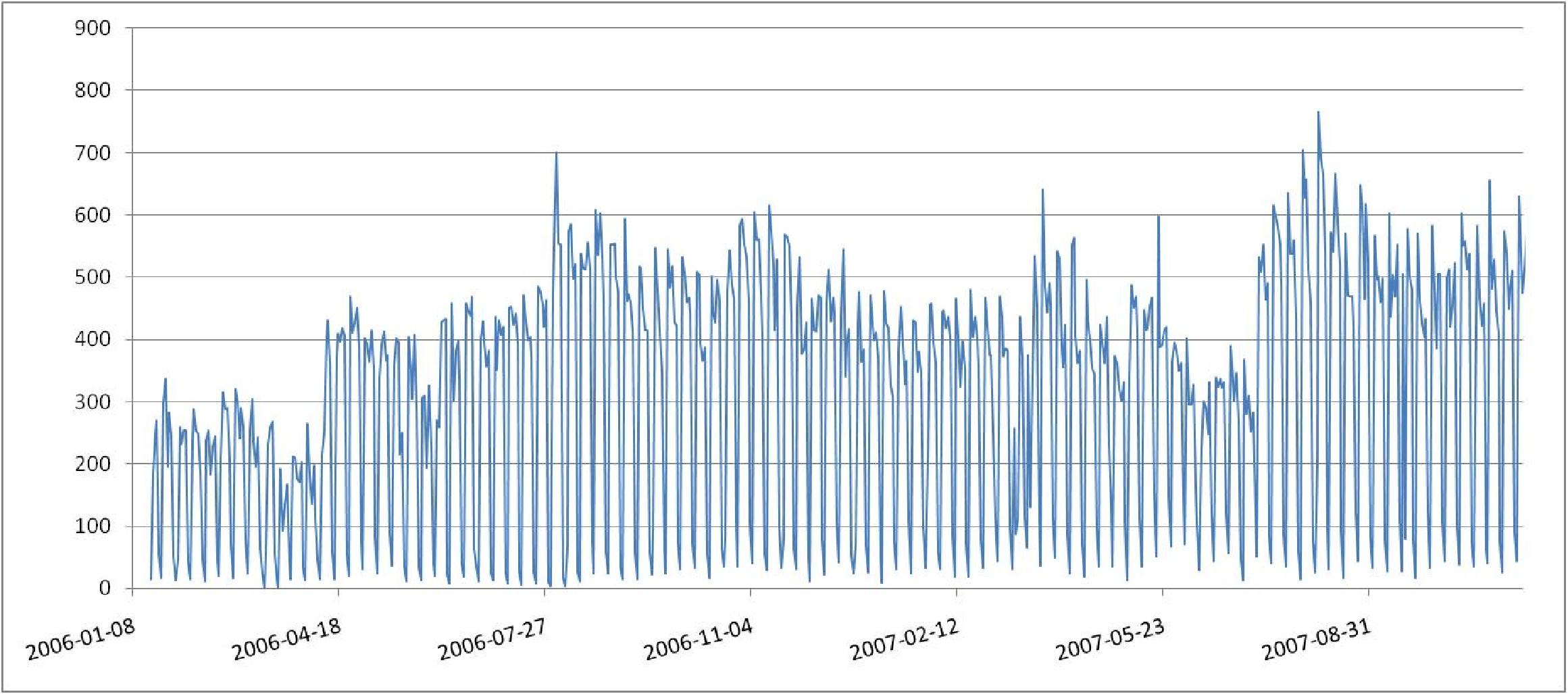} \\
\end{array}$
\caption{Four call center series, out of 21.}
\end{figure}

In our data set the series had on average 760 points, ranging from 383 for the shortest to 826 for the longest. Four typical series are shown in Figure \ref{series}.

We used a set $\mathcal{D}$ of selected dates $m < n$ and, for each method and each time series $(y_1, \ldots, y_n)$ (here, $n=760$ on average), we trained the models on the pruned series $(y_1, \ldots, y_m)$ and predicted the $\tau$-quantile at time $m+1$.  The set  $\mathcal{D}$ is composed of  $91$ dates, so that all quality criteria used to measure the proximity between the predicted quantiles and the observed values $y_{m+1}$ were computed using $91 \times 21 = 1911$ points. The 21 times series and the set $\mathcal D$ are available at the address  \texttt{http://www.lsta.upmc.fr/doct/patra/}\,.\\

In a first series of experiments, we let the methods predict the $\tau$-quantiles at the 1911 dates for $\tau \in \{0.1, 0.5,0.9\}$. We compared the performances of our expert-based strategy, denoted hereafter by $\texttt{QuantileExpertMixture}_\tau$, with those of $\texttt{QAR(p)}_\tau$, a $\tau$-quantile linear autoregressive model of order $p$. This quantile prediction model, which is described in \cite{KOE1}, also uses the pinball criterion to fit its parameters. The implementation we used solves the minimization problem with an Iterative Re-weighted Least Square algorithm (IRLS), see for instance Street, Caroll and Ruppert \cite{CAR2}. Following Takeuchi, Le, Sears and Smola \cite{TAK1}, we used two criteria to measure the quality of the overall set of quantile forecastings. First, we evaluated the expected risk with respect to the pinball function $\rho_\tau$, referred to as \small {\sc PinBall Loss} in the sequel. 
Secondly we calculated \small {\sc Ramp Loss}, the empirical fraction of quantile estimates which exceed the observed values $y_{m+1}$. Ideally, the value of \small {\sc Ramp Loss} should be close to $1 - \tau$. \\

Tables \ref{tableau1}-\ref{tableau3} show the $\texttt{QuantileExpertMixture}_\tau$ and $\texttt{QAR(p)}_{\tau}$ results at the selected dates $\mathcal D$ of the call center series. The latter algorithm was benchmarked for each order $p$ in $\{1,\hdots, 10\}$, but we reported only the most accurate order $p = 7$. The best results with respect to each criterion are shown in bold. We see that both methods perform roughly similarly, with eventually a slight advantage for the autoregressive strategy for $\tau =0.1$ whereas $\texttt{QuantileExpertMixture}_{\tau}$ does better for $\tau = 0.9$.

\begin{table}[!hf]
\centering
\begin{tabular}{|c||c|c|}
\hline
      \texttt{Method} & \small{\sc PinBall Loss (0.1)} & \small {\sc Ramp Loss} \\
      \hline
      \hline
      $\texttt{QuantileExpertMixture}_{0.1}$ & 13.71 & 0.80 \\
      $\texttt{QAR(7)}_{0.1}$ & \textbf{13.22} & \textbf{0.88} \\
      \hline
\end{tabular}
\caption{Quantile forecastings with $\tau = 0.1$.}
\label{tableau1}
\end{table}

\begin{table}[!hf]
\centering
\begin{tabular}{|c||c|c|}
\hline
\texttt{Method} & \small{\sc PinBall Loss (0.5)} & \small {\sc Ramp Loss} \\
      \hline
      \hline
      $\texttt{QuantileExpertMixture}_{0.5} $ & \textbf{24.05} & 0.42\\
      $\texttt{QAR(7)}_{0.5}$ & 29.157 & \textbf{0.47} \\
      \hline
\end{tabular}
\caption{Quantile forecastings with $\tau = 0.5$.}
\label{tableau2}
\end{table}

\begin{table}[!hf]
\centering
\begin{tabular}{|c||c|c|}
\hline
      \texttt{Method} & \small{\sc PinBall Loss (0.9)} & \small {\sc Ramp Loss}\\
      \hline
      \hline
      $\texttt{QuantileExpertMixture}_{0.9} $ & \textbf{12.27} & \textbf{0.07}\\
      $\texttt{QAR(7)}_{0.9}$ & 19.31 & \textbf{0.07}\\
      \hline
\end{tabular}
\caption{Quantile forecastings with $\tau = 0.9$.}
\label{tableau3}
\end{table}

\newpage

Median-based predictors are well known for their robustness while predicting individual values for time series, see for instance Hall, Peng and Yao \cite{HAL1}. Therefore, in a second series of experiments,  we fixed $\tau=0.5$ and focused on the problem of predicting future outcomes of the series. We decided to compare the results of $\texttt{QuantileExpertMixture}_{0.5} $ with those of 6 concurrent predictive procedures:
\begin{itemize}
\item $\texttt{MA}$ denotes the simple moving average model.
\item $\texttt{AR(p)}$ is a linear autoregressive model of order $p$, with parameters computed with respect to the usual least square criterion.
\item $\texttt{QAR(p)}$ is the $\tau$-quantile linear autoregressive model of order $p$ described earlier.
\item $\texttt{DayOfTheWeekMA}$ is a naive model, which applies moving averages on the days of the week, that is a moving average on the Sundays, Mondays, and so on.
\item $\texttt{MeanExpertMixture}$ is an online prediction algorithm described in \cite{BIA1}. It is based on conditional mean estimation and close in spirit to the strategy $\texttt{QuantileExpertMixture}_{0.5}$.
\item And finally, we let $\texttt{HoltWinters}$ be the well-known procedure which performs exponential smoothing on three components of the series, namely Level, Trend and Seasonality. For a thorough presentation of $\texttt{HoltWinters}$ techniques we refer the reader to Madrikakis, Whellwright and Hyndman \cite{MAD1}.
\end{itemize}

Accuracy of all forecasting methods were measured using the Average Absolute Error (\small{\sc Avg Abs Error}, which is proportional to the pinball error since $\tau=0.5$),  Average Squared Error (\small{\sc Avg Sqr Error}), and the unstable but widely spread criterion Mean Average Percentage Error (\small{\sc MAPE}, see \cite{MAD1} for definition and discussion). We also reported the figure \small{\sc Abs Std Dev} which corresponds to the empirical standard deviation of the differences $|y^F_t - y^R_t|$, where $y^F_t$ stands for the forecasted value while $y^R_t$ stands for the observed value of the time series at time $t$. $\texttt{AR(p)}$ and $\texttt{QAR(p)}$ algorithms were run for each order $p$ in $\{1,\hdots, 10\}$, but we reported only the most accurate orders. \\

\begin{table}[!hf]
\centering
\begin{footnotesize}
\begin{tabular}{|c||c|c|c|c|}
\hline
       \texttt{Method} & \sc Avg Abs Error & \sc Avg Sqr Error & \sc MAPE (\%) & \sc Abs Std Dev\\
      \hline
      \hline
      $\texttt{MA}$ & 179.0 & 62448 & 52.0 & 174.8\\
      $\texttt{AR(7)}$ & 65.8 & 9738 & 31.6 & 73.5 \\
      $\texttt{QAR(8)}_{0.5}$ & 57.8 & 9594 & 24.9 & 79.2 \\
      $\texttt{DayOfTheWeekMA}$ & 54.1 & 7183 & 22.8 & 64.7\\
       $\texttt{QuantileExpertMixture}_{0.5}$ & \textbf{48.1} & \textbf{5731} & 21.6 &  \textbf{58.4}\\
      $\texttt{MeanExpertMixture}$ & 52.4 & 6536 & 22.3 & 61.6 \\
      $\texttt{HoltWinters}$ & 49.8 & 6025 & \textbf{21.5} & 59.5\\
      \hline
      \end{tabular}
\end{footnotesize}
\caption{Future outcomes forecastings.}
\label{grostableau}
\end{table}

We see via Table \ref{grostableau} that the nearest neighbor strategy presented here outperforms all other methods in terms of Average Absolute Error.  Interestingly, this forecasting procedure also provides the best results with respect to the Average Squared Error criterion. 
This is remarkable, since $\texttt{QuantileExpertMixture}_{0.5}$ does not rely on a squared error criterion, contrary to $\texttt{MeanExpertMixture}$. The same comment applies to $\texttt{QAR(8)}_{0.5}$ and $\texttt{AR(7)}$. In terms of the Mean Average Percentage Error, the present method and \texttt{HoltWinters} procedure provide good and broadly similar results.
\section{Proofs}
\subsection{Proof of Theorem \ref{thm:main}}
The following lemmas will be essential in the proof of Theorem \ref{thm:main}. The first one is known as Breiman's generalised
ergodic theorem (Breiman \cite{BRE1}).
\begin{lem}
\label{thm:Breiman}
 Let $Z=\{Z_n\}_{-\infty}^{\infty}$ be a stationary and
ergodic process. For each positive integer $t$, let $T^t$ denote the
left shift operator, shifting any sequence of real numbers $\{\hdots, z_{-1},
z_0,z_1, \hdots\}$ by $t$ digits to the left. Let $\{f_t\}_{t=1}^{\infty}$
be a sequence of real-valued functions such that $\lim_{t \to
\infty}f_t(Z) =f(Z)$ a.s. for some function $f$. Suppose
that $\mathbb E[\sup_t |f_t(Z)|]<\infty$. Then
$$\lim_{n \to \infty} \frac{1}{n} \sum_{t=1}^n f_t(T^tZ)=\mathbb E \left[ f(Z)\right]\quad a.s.$$
\end{lem}
Lemma \ref{lem:Ineq} below is due to Gy\"{o}rfi and Ottucs\'{a}k \cite{GYO5}. These authors proved the inequality for any cumulative normalized loss of form $L_n(h)= \frac{1}{n}\sum_{t=1}^n \ell_t(h)$, where $\ell_t(h) =\ell_t(h_t,Y_t)$ is convex in its first argument, what is the case for the function $\ell_t(h_t,Y_t)=\rhotau ( Y_t - h_t(Y_1^{t-1}))$.

\begin{lem}
\label{lem:Ineq} Let $g=\{g_n\}_{n=1}^{\infty}$ be the nearest neighbor quantile prediction strategy defined in (\ref{QPS}). Then, for every $n \ge 1$, a.s.,
\begin{align*}
L_n(g) \leq &\inf_{k,\ell}\left(L_n(h_n^{(k,\ell)}) - \frac{2\ln b_{k,\ell} }{n\eta_{n+1}} \right) \\
& \quad + \frac{1}{2n} \sum_{t=1}^{n}{\eta_{t} \sum_{k,\ell = 1}^{\infty}{p_{k,\ell,n}{\left[\rhotau \left(Y_t - h_t^{(k,\ell)}(Y_1^{t-1})\right)\right]^2}}}.
\end{align*}
\end{lem}

\begin{lem}
\label{lem:tech}
Let $x,y \in \mathbb R$ and $\ell \in \mathbb{N}$. Then
\begin{enumerate}
\item $\rhotau(x) \leq |x|.$
\item $\rhotau (x + y) \leq \rhotau (x) +  \rhotau (y).$
\item $\rhotau\left(T_\ell(x) -  T_\ell(y)\right) \leq \rhotau (x - y).$
\end{enumerate}
\end{lem}
{\bf Proof of Lemma \ref{lem:tech}}\quad Let $x,y \in \mathbb R$ and $\ell \in \mathbb{N}$.
\begin{enumerate}
\item We have
$$\left |\rhotau(x) \right | = \left| x (\tau - \mathbf 1_{[x \leq 0]}) \right| = |x| \left|\tau - \mathbf 1_{[x \leq 0]} \right|\leq |x|.$$
\item Clearly,
\begin{align*}
&\rhotau(x + y) \leq \rhotau(x) + \rhotau(y) \\
\Longleftrightarrow \quad &  x\mathbf 1_{[x \leq 0]} + y \mathbf 1_{[y \leq 0]} \leq x \mathbf 1_{[x+y \leq 0]} + y\mathbf 1_{[x+y \leq 0]}.
\end{align*}
The conclusion follows by examining the different positions of $x$ and $y$ with respect to $0$.
\item  If $x > \ell$ and $|y| \leq \ell$, then
\begin{align*}
\rhotau\left(T_\ell(x) -  T_\ell(y)\right)& = \rhotau(\ell - y) \\
&= (\ell -y )(\tau - \mathbf 1_{[\ell - y \leq 0]}) \\
&=  (\ell -y )\tau \\
& \leq (x - y) \tau \\
& = (x - y)(\tau - \mathbf 1_{[x -y \leq 0]})\\
&= \rhotau(x-y).
\end{align*}
Similarly, if $x < -\ell$ and $|y| \leq \ell$, then
\begin{align*}
\rhotau\left(T_\ell(x) -  T_\ell(y)\right)& = \rhotau(-\ell - y) \\
& =(-\ell -y)(\tau - \mathbf1 _{[-\ell - y \leq 0]}) \\
& = (-\ell -y)(\tau-1)\\
&  \leq (x -y)(\tau - 1)\\
& =(x -y)(\tau - \mathbf 1_{[x-y \leq 0]})\\
&= \rhotau(x-y).
\end{align*}
All the other cases are similar and left to the reader.
\end{enumerate}
\begin{flushright}
$\square$
\end{flushright}
Recall that a sequence $\{\mu_n\}_{n=1}^{\infty}$ of probability measures on $\mathbb R$ is defined to converge weakly to the probability measure $\mu_{\infty}$ if for every bounded, continuous real function $f$,
$$\int f \mbox{d}\mu_n \to \int f \mbox{d} \mu_\infty \quad \mbox{as } n \to \infty.$$
Recall also that the sequence $\{\mu_n\}_{n=1}^{\infty}$ is said to be uniformly integrable if
$$\lim_{\alpha \to \infty} \sup_{n\geq 1} \int_{|x| \geq \alpha }{ |x| \mbox{d}\mu_n(x)} = 0.$$
Moreover, if
$$ \sup_{n \geq 1}{\int|x|^{1+\varepsilon}  \mbox{d}\mu_n(x) } < \infty$$
for some positive $\varepsilon$, then the sequence  $\{\mu_n\}_{n=1}^{\infty}$ is uniformly integrable (Billingsley \cite{BIL1}).\\

The next lemma may be summarized by saying that if a sequence of probability measures converges in terms of weak convergence topology, then the associated quantile sequence will converge too.
\begin{lem}
\label{lem:pat}
Let $\{\mu_n\}_{n=1}^{\infty}$ be a uniformly integrable sequence of real probability measures, and let $\mu_{\infty}$ be a probability measure with (strictly) increasing distribution function. Suppose that
$\{\mu_n\}_{n=1}^{\infty}$ converges weakly to $\mu_{\infty}$. Then, for all $\tau \in (0,1)$,
$$q_{\tau,n}  \to  q_{\tau,\infty} \quad \mbox{as } n \to \infty,$$
where $q_{\tau,n} \in \mathcal Q_{\tau}(\mu_n)$ for all $n \geq 1$ and $\{q_{\tau,\infty}\} = \mathcal Q_{\tau}(\mu_{\infty})$.
\end{lem}
{\bf Proof of Lemma \ref{lem:pat}}\quad Since $\{\mu_n\}_{n=1}^{\infty}$ converges weakly to $\mu_{\infty}$, it is a tight sequence. Consequently, there is a compact set, say $[-M,M]$, such that $\mu_n(\mathbb R \setminus [-M,M]) < \min(\tau, 1 - \tau )$. This implies $q_{\tau,n} \in [-M,M]$ for all $n\geq 1$. Consequently, it will be enough to prove that any consistent subsequence of $\{q_{\tau, n}\}_{n=1}^{\infty}$ converges towards $q_{\tau,\infty}$.\\

\begin{figure}[!hf]
\begin{center}
\label{rhotaudec}
\includegraphics[scale = 0.75]{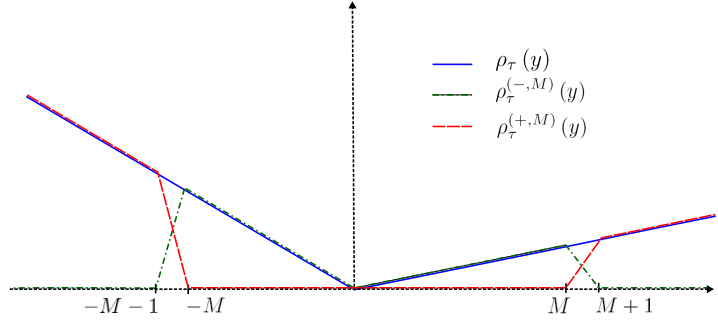}
\caption{Illustration of the decomposition $\rho_\tau=\rho_\tau^{(+,M)} + \rho_\tau^{(-,M)}$. }
\end{center}
\end{figure}

Using a slight abuse of notation, we still denote by $\{q_{\tau,n}\}_{n=1}^{\infty}$ a consistent subsequence of the original sequence, and let $q_{\tau, \star}$ be such that $\lim_{n \to \infty} q_{\tau, n} =q_{\tau,\star}$. Using the assumption on the distribution function of $\mu_{\infty}$, we know by Lemma \ref{tilt} that $q_{\tau,\infty}$ is the unique minimizer of problem (\ref{wep}). Therefore, to show that $q_{\tau,\star}=q_{\tau,\infty}$, it suffices to prove that, for any $q \in \mathbb R$,
$$\mathbb{E}_{\mu_\infty}\left[ \rhotau(Y -q)\right] \geq \mathbb{E}_{\mu_\infty}\left[ \rhotau(Y -q_{\tau,\star})\right].$$
Fix $q \in \mathbb R$. We first prove that
\begin{equation}\label{eq:pat1}
\mathbb{E}_{\mu_n}\left[\rhotau(Y -q)\right] \to \mathbb{E}_{\mu_\infty}\left[ \rhotau(Y -q)\right] \quad \mbox{as } n \to \infty.
\end{equation}
To see this, for $M >0$ and all $y\in \mathbb{R}$, set

\[
 \rho_\tau^{(+,M)}(y) = \left \{ \begin{array}{ll}
       0 & \mbox{if $|y| < M$;} \\
       \rho_\tau(y) & \mbox{if $|y| > M+1$;}\\
       \rho_\tau(M+1)(y-M) & \mbox{if $y \in [M, M+1]$;}\\
       \rho_\tau(-M-1)(y+M) & \mbox{if $y \in [-M-1, -M]$.}
                \end{array}   \right.
\]
The function $\rho_\tau^{(+,M)}$ is continuous and, for all $z \in \mathbb{R}$, satisfies the inequality $\rho_\tau^{(+,M)}(z) \leq \rho_\tau(z)\mathbf 1_{[|z| > M]}$. In the sequel, we will denote by $\rho_\tau^{(-,M)}$ the bounded and continuous map $\rho_\tau - \rho_\tau^{(+,M)}$. The decomposition $\rho_\tau = \rho_\tau^{(+,M)} + \rho_\tau^{(-,M)}$ is illustrated in Figure \ref{rhotaudec}.

Next, fix $\varepsilon > 0$ and choose $M$ large enough to ensure
$$\sup_{n \geq 1}\left(\esp_{\mu_n}\left[|Y-q| \mathbf 1_{[|Y -q| > M]}\right]\right) + \esp_{\mu_\infty}\left[ |Y-q| \mathbf 1_{[|Y -q| > M]} \right] <\varepsilon/2.$$
Choose also $n$ sufficiently large to have
$$\left|\esp_{\mu_n}\left[\rhotau^{(-,M)}(Y-q)\right] - \esp_{\mu_\infty}\left[\rhotau^{(-,M)}(Y-q)\right]\right| < \varepsilon/2.$$
Write
\begin{align*}
& \left| \esp_{\mu_n}\left[\rhotau(Y-q)\right] - \esp_{\mu_\infty} \left[\rhotau(Y-q)\right]\right| \\
& \quad =\Big|\esp_{\mu_n}\left[\rhotau^{(+,M)}(Y-q)\right] + \esp_{\mu_n}\left[\rhotau^{(-,M)}(Y-q)\right] \\
& \quad \qquad - \esp_{\mu_\infty}\left[\rhotau^{(+,M)}(Y-q)\right] - \esp_{\mu_\infty}\left[\rhotau^{(-,M)}(Y-q)\right] \Big|.\\
\end{align*}
Thus
\begin{align*}
& \left| \esp_{\mu_n}\left[\rhotau(Y-q)\right] - \esp_{\mu_\infty} \left[\rhotau(Y-q)\right]\right| \\
& \quad  \leq \left| \esp_{\mu_n}\left[\rhotau^{(-,M)}(Y-q)\right] - \esp_{\mu_\infty}\left[\rhotau^{(-,M)}(Y-q)\right]\right| \\
& \quad \qquad + \left|\esp_{\mu_n}\left[\rhotau^{(+,M)}(Y-q)\right] - \esp_{\mu_\infty}\left[\rhotau^{(+,M)}(Y-q)\right] \right|\\
& \quad \leq \left|\esp_{\mu_n}\left[\rhotau^{(-,M)}(Y-q)\right] - \esp_{\mu_\infty}\left[\rhotau^{(-,M)}(Y-q)\right]\right| \\
& \quad \qquad + \left|\esp_{\mu_n}\left[\rhotau(Y-q)\mathbf 1_{[|Y-q| > M ]}\right]\right| + \left|\esp_{\mu_\infty}\left[\rhotau(Y-q)\mathbf 1_{[|Y-q| > M]}\right]\right|\\
& \quad \leq \left|\esp_{\mu_n}\left[\rhotau^{(-,M)}(Y-q)\right] - \esp_{\mu_\infty}\left[\rhotau^{(-,M)}(Y-q)\right]\right| \\
& \quad \qquad + \sup_n\esp_{\mu_n}\left[|Y-q|\mathbf 1_{[|Y-q| > M ]}\right]+ \esp_{\mu_\infty}\left[|Y-q|\mathbf 1_{[|Y-q| > M ]}\right]\\
& \quad \leq \varepsilon.
\end{align*}
for all large enough $n$. This shows (\ref{eq:pat1}).\\

Next, using the fact that the function $\rhotau$ is uniformly continuous, we may write, for sufficiently large $n$ and all $y\in \mathbb R$,
\begin{equation}
\label{eq:pat2}
 \rhotau(y -q_{\tau,n}) \geq \rhotau(y - q_{\tau,\star}) - \varepsilon.
 \end{equation}
Therefore, for all large enough $n$,
\begin{align*}
\mathbb{E}_{\mu_\infty}\left[ \rhotau(Y-q)\right] &\geq \mathbb{E}_{\mu_n}\left[ \rhotau(Y-q)\right] - \varepsilon\\
& \quad (\mbox{by identity }(\ref{eq:pat1}))\\
&\geq \mathbb{E}_{\mu_n}\left[ \rhotau(Y-q_{\tau,n})\right] - \varepsilon \\
&\geq \mathbb{E}_{\mu_n}\left[ \rhotau(Y-q_{\tau,\star}\right)]- 2\varepsilon\\
& \quad (\mbox{by inequality}(\ref{eq:pat2}))\\
&\geq \mathbb{E}_{\mu_\infty}\left[ \rhotau(Y-q_{\tau,\star})\right] - 3\varepsilon\\
& \quad (\mbox{by identity }(\ref{eq:pat1})).\\
\end{align*}
Letting $\varepsilon \to 0$ leads to the desired result.
\begin{flushright}
$\square$
\end{flushright}
We are now in a position to prove Theorem \ref{thm:main}.\\

Because of inequality (\ref{prop:alg}) it is enough to show that
\[
\limsup_{n\to \infty}L_n(g)\le L^{\star} \quad \mbox{a.s.}
\]
With this in mind, we first provide an upper bound on the first term of the right hand side of the inequality in Lemma \ref{lem:Ineq}. We have
\begin{align*}
&\limsup_{n \rightarrow \infty} \inf_{k,\ell} \left( L_n\left(h_n^{(k,\ell)} -\frac{2\ln b_{k,\ell}}{n\eta_{n+1}} \right)\right)\\
& \quad \leq \inf_{k,\ell} \left( \limsup_{n \to \infty} L_n\left(h_n^{(k,\ell)} -\frac{2\ln b_{k,\ell}}{n\eta_{n+1}} \right) \right)\\
& \quad \leq \inf_{k,\ell} \left( \limsup_{n \to \infty} L_n\left( h_n^{(k,\ell)}\right) \right).
\end{align*}
To evaluate $\limsup_{n \to \infty} L_n ( h_n^{(k,\ell)})$, we investigate the performance of the expert $h_n^{(k,\ell)}$ on the stationary and ergodic sequence $Y_0, Y_{-1},Y_{-2}, \hdots$ Fix $p_\ell \in (0,1)$, $\textbf{s} \in \mathbb R^k$, and set $\tilde{\ell}=\lfloor p_\ell j\rfloor$, where $j$ is a positive integer.\\

For $j> k + \tilde{\ell} +1$, introduce the set
\begin{align*}
 \tilde{J}_{j,\textbf{s}}^{(k,\tilde{\ell})} = &\left\{  -j + k +1 \leq i \leq 0 : Y_{i-k}^{i-1} \text{ is among the } \tilde{\ell}\text{-NN of } \textbf{s} \right.\\
 & \quad \left.\text{ in } Y_{-k}^{-1}, \ldots , Y_{-j+1}^{-j+k} \right\}.
 \end{align*}
For any real number $a$, we denote by $\delta_{a}$ the Dirac (point) measure at $a$. Let the random measure $\prob_{j,\textbf{s}}^{(k,\ell)}$ be defined by
$$\prob_{j,\textbf{s}}^{(k,\ell)} = \frac{1}{\left|\tilde{J}_{j, \textbf{s}}^{(k,\tilde{\ell})}\right|}{\sum_{i \in \tilde{J}_{j,\textbf{s}}^{(k,\tilde{\ell})}}\delta_{Y_i}}.$$

Take an arbitrary radius $r_{k,\ell}(\textbf{s})$ such that
\begin{equation*}
\prob\left[ \| Y_{-k}^{-1}- \textbf{s} \| \leq r_{k,\ell}(\textbf{s}) \right]=p_\ell.
\end{equation*}
A straightforward adaptation of an argument in Theorem 3.1 of \cite{GYO8} shows that
$$\prob_{j,\textbf{s}}^{(k,\ell)} \underset{j \to \infty}{\to} \prob_{Y_0 \,|\, \|Y_{-k}^{-1} - \textbf{s} \| \leq r_{k,\ell}(\textbf{s})} \triangleq \prob^{(k,\ell)}_{\infty,\textbf{s}}$$
almost surely in terms of weak convergence. Moreover, by a double application of the ergodic theorem (see for instance \cite{BIA1}),
$$\int{ y^2 \mbox{d}\prob_{j,\textbf{s}}^{(k,\ell)}(y)} \underset{j \to \infty}{\to} \int{ y^2 \mbox{d}\prob^{(k,\ell)}_{\infty,\textbf{s}}(y)} \quad \mbox{a.s.}$$
Thus
$$\sup_{j \geq 0}{\int{y^2 \mbox{d}\prob_{j,\textbf{s}}^{(k,\ell)}(y)}} < \infty  \quad \mbox{\text{ a.s.}},$$
and, consequently, the sequence $\{\prob_{j,\textbf{s}}^{(k,\ell)}\}_{j=1}^{\infty}$ is uniformly integrable. \\

By assumption $(H3)$ the distribution function of the measure $\prob_{Y_0|Y_{-\infty}^{-1}}$ is a.s. increasing. We also have $\sigma\left(\|Y_{-k}^{-1} - \textbf{s} \| \leq r_{k,\ell}(\textbf{s})\right) \subset \sigma\left(Y_{-\infty}^{-1}\right)$ where $\sigma(X)$ denotes the sigma algebra generated by the random variable $X$. Thus the distribution function of $\prob^{(k,\ell)}_{\infty,\textbf{s}}= \prob_{Y_0 \,|\, \|Y_{-k}^{-1} - \textbf{s} \| \leq r_{k,\ell}(\textbf{s})}$ is a.s. increasing, too. Hence, letting
$$q_{\tau,j}^{(k,\ell)}(Y_{-j+1}^{-1},\textbf{s}) \in \mathcal Q_\tau (\prob_{j,\textbf{s}}^{(k,\ell)})\quad \mbox{and} \quad \left\{q_{\tau,\infty}^{(k,\ell)}(\textbf{s})\right\} = \mathcal Q_\tau(\prob^{(k,\ell)}_{\infty,\textbf{s}}),$$
we may apply Lemma \ref{lem:pat}, and obtain
$$q_{\tau,j}^{(k,\ell)} (Y_{-j+1}^{-1},\textbf{s})\underset{j \to \infty}{\to} q_{\tau,\infty}^{(k,\ell)}\left(\textbf{s}\right) \quad \mbox{\text{ a.s.}}$$
Consequently, for any $y_0 \in \mathbb R$,
$$\rho_{\tau}\left(y_0-T_{\min(j^{\delta},\ell)}\left(q_{\tau,j}^{(k,\ell)} (Y_{-j+1}^{-1},\textbf{s} ) \right) \right) \underset{j \to \infty}{\to} \rho_{\tau}\left( y_0 - T_\ell \left(q_{\tau,\infty}^{(k,\ell)}(\textbf{s})\right)\right) \quad \mbox{\text{ a.s.}}$$
Since $y_0$ and $\textbf{s}$ are arbitrary, we are led to
\begin{equation}
\label{eq:conv}
\rho_{\tau}\left(Y_0-T_{\min(j^{\delta},\ell)}\left(q_{\tau,j}^{(k,\ell)} (Y_{-j+1}^{-1},Y_{-k}^{-1}) \right) \right) \underset{j \to \infty}{\to} \rho_{\tau}\left( Y_0 - T_\ell \left(q_{\tau,\infty}^{(k,\ell)}(Y_{-k}^{-1})\right)\right) \quad \mbox{\text{ a.s.}}
\end{equation}
For $y= (\hdots, y_{-1}, y_0,y_1, \hdots)$, set
\begin{align*}
f_j (y) &\triangleq \rho_{\tau}\left (y_0-h_j^{(k,\ell)} (y_{-j + 1 }^{-1} )\right) \\
&= \rho_{\tau}\left(y_0-T_{\min(j^\delta,\ell)}\left( q_{\tau,j}^{(k,\ell)} ( y_{-j + 1}^{-1},y_{-k}^{-1})\right)\right).
\end{align*}
Clearly,
\begin{align*}
\left|f_j ( Y )\right| &= \left|\rhotau\left(Y_0 - h_j^{(k,\ell)}(Y_{-j+1}^{-1}) \right)\right|\\
& \leq \left|Y_0 - T_{\min(j^\delta,\ell)} ( Y_{-j+1}^{-1})\right| \\
& \quad (\mbox{by statement 1. of Lemma \ref{lem:tech}})\\
& \leq |Y_0| + \left|T_{\min(j^\delta,\ell)}(Y_{-j+1}^{-1})\right| \\
& \leq |Y_0|+ \ell,
\end{align*}
and thus $\esp [\sup_{j} |f_j(Y) |] < \infty$. By identity (\ref{eq:conv}),
$$f_j (Y) \underset{j \to \infty}{\to} \rho_{\tau}\left( Y_0 - T_\ell(q_{\tau,\infty}^{(k,\ell)}(Y_{-k}^{-1})\right)\quad \mbox{a.s.}$$
Consequently,
Lemma \ref{thm:Breiman} yields
$$L_n ( h_n^{(k,\ell)}) \underset{n \to \infty}{\to} \mathbb{E}\left[ \rhotau\left( Y_0 - T_\ell\left(q_{\tau,\infty}^{(k,\ell)}(Y_{-k}^{-1})\right)\right) \right].$$
To lighten notation a bit, we set
$$ \varepsilon_{k,\ell} \triangleq \mathbb{E}\left[ \rhotau\left( Y_0 - T_\ell\left(q_{\tau,\infty}^{(k,\ell)}(Y_{-k}^{-1})\right)\right) \right]$$
and proceed now to prove that $\lim_{k \to \infty} \lim_{\ell \to \infty} \varepsilon_{k,\ell}\leq L^{\star}$.\\

We have, a.s., in terms of weak convergence,
$$ \prob^{(k,\ell)}_{\infty,Y_{-k}^{-1}} \underset{\ell \to \infty }{\to} \prob_{Y_0 \,|\, Y_{-k}^{-1}}$$
(see for instance Theorem 3.1 in \cite{GYO8}).  Next, with a slight modification of techniques of Theorem 2.2 in \cite{BIA1},
$$\int{ y^2 \mbox{d} \prob^{(k,\ell)}_{\infty,Y_{-k}^{-1}}(y)} \underset{\ell \to \infty}{\to} \int{ y^2 \mbox{d}\prob_{Y_0 \,|\, Y_{-k}^{-1}}(y)} \quad \mbox{\text{ a.s.}},$$
which leads to
$$\sup_{\ell \geq 0}{\int{y^2 \mbox{d}\prob_{\infty,Y_{-k}^{-1}}^{(k,\ell)}(y)}} < \infty  \quad \mbox{a.s.}$$
Moreover, by assumption $(H3)$, the distribution  function of $\prob_{Y_0\,|\, Y_{-k}^{-1}}$ is a.s. increasing. Thus, setting
$$\left\{q_{\tau,\infty}^{(k,\ell)}(Y_{-k}^{-1}) \right\} =\mathcal Q_{\tau} (\prob^{(k,\ell)}_{\infty,Y_{-k}^{-1}}) \quad \mbox{and} \left\{q_\tau^{(k)}(Y_{-k}^{-1})\right\}  =  \mathcal Q_\tau(\prob_{Y_0 \,|\, Y_{-k}^{-1}})$$
and applying Lemma \ref{lem:pat} yields
\begin{equation*}
\label{eq:convquantile}
 q_{\tau,\infty}^{(k,\ell)}(Y_{-k}^{-1})\underset{\ell \to \infty}{\to} q_\tau^{(k)}(Y_{-k}^{-1}) \quad \mbox{a.s.}
\end{equation*}
Consequently,
$$ \rho_{\tau}\left(Y_0 - T_\ell \left(q_{\tau,\infty}^{(k,\ell)}(Y_{-k}^{-1})\right)\right) \underset{\ell \to \infty}{\to} \rho_{\tau}\left(Y_0 - q_\tau^{(k)}(Y_{-k}^{-1})\right) \quad \mbox{}\text{a.s.}$$
It turns out that the above convergence also holds in mean. To see this, note first that
\begin{align*}
& \rhotau\left(Y_0 - T_\ell\left(q_{\tau,\infty}^{(k,\ell)}( Y_{-k}^{-1}) \right)\right) \\
& \quad = \rhotau\left(Y_0 - T_\ell (Y_0) + T_\ell (Y_0) - T_\ell\left(q_{\tau,\infty}^{(k,\ell)}( Y_{-k}^{-1}) \right)\right) \\
& \quad \leq \rhotau\left(Y_0 - T_\ell(Y_0)\right) + \rhotau\left(T_\ell (Y_0) - T_\ell\left(q_{\tau,\infty}^{(k,\ell)}( Y_{-k}^{-1})\right)\right)\\
&\qquad  (\mbox{by statement 2. of Lemma \ref{lem:tech}}) \\
&\quad \leq 2|Y_0| + \rhotau\left(Y_0- q_{\tau,\infty}^{(k,\ell)}( Y_{-k}^{-1}) \right) \quad \mbox{\text{ a.s.}}\\
& \qquad (\mbox{by statement 3. of Lemma \ref{lem:tech}}).
\end{align*}
Thus
\begin{align*}
&\esp \left[\left(\rhotau\left(Y_0 - T_\ell \left (q_{\tau,\infty}^{(k,\ell)}( Y_{-k}^{-1})\right)\right) \right)^2\right] \\
& \quad \leq \esp \left[\left( 2|Y_0| + \rhotau \left(Y_0- q_{\tau,\infty}^{(k,\ell)}( Y_{-k}^{-1})\right)\right)^2\right] \\
& \quad \leq 8\esp \left[Y_0^2\right] + 2 \esp\left [\left(\rhotau\left(Y_0- q_{\tau,\infty}^{(k,\ell)}( Y_{-k}^{-1}) \right)\right)^2\right].
\end{align*}
In addition,
\begin{align*}
&\sup_{\ell \geq 1}{\esp\left[\left(\rhotau\left(Y_0- q_{\tau,\infty}^{(k,\ell)}( Y_{-k}^{-1})\right)\right)^2\right]} \\
& \quad = \sup_{\ell \geq 1}{\esp\left[\left( \min_{q(.)} \esp_{\prob^{(k,\ell)}_{\infty,Y_{-k}^{-1}}} \rhotau\left(Y_0- q(Y_{-k}^{-1})\right)\right)^2\right]} \\
& \quad \leq \esp \left[\left(\rhotau (Y_0 )\right)^2\right]  \\
& \qquad \mbox{(by Jensen's inequality)}\\
& \quad \leq \esp \left[Y_0 ^2\right]  < \infty.
\end{align*}
This implies
 $$\esp \left[\left(\rhotau\left(Y_0 - T_\ell \left (q_{\tau,\infty}^{(k,\ell)}( Y_{-k}^{-1})\right)\right) \right)^2\right] < \infty,$$
 i.e., the sequence is uniformly integrable. Thus we obtain, as desired,
 $$ \lim_{\ell \to \infty} \mathbb{E}\left[\rho_{\tau}\left(Y_0 - T_\ell\left(q_{\tau,\infty}^{(k,\ell)}( Y_{-k}^{-1})\right)\right) \right] = \mathbb{E}\left[\rho_{\tau}\left(Y_0 - q_{\tau}^{(k)}(Y_{-k}^{-1})\right)\right].$$
 Putting all pieces together,
\begin{align*}
\lim_{\ell \to \infty}{\varepsilon_{k,\ell}} & = \lim_{\ell \to \infty} \mathbb{E}\left[\rho_{\tau}\left(Y_0 - T_\ell\left(q_{\tau,\infty}^{(k,\ell)}( Y_{-k}^{-1})\right)\right) \right] \\
& = \mathbb{E}\left[\rho_{\tau}\left(Y_0 - q_{\tau}^{(k)}(Y_{-k}^{-1})\right)\right] \\
& \triangleq \varepsilon_k^{\star}.
\end{align*}
It remains to prove that $\lim_{k \to \infty} \varepsilon_k^{\star} = L^{\star}$. To this aim, for all $k \geq 1$, let $Z_k$ be the $\sigma ( Y_{-k}^{-1})$-measurable random variable defined by
$$ Z_k = \rhotau \left(Y_0 - q_\tau^{(k)}(Y_{-k}^{-1}) \right) = \min_{q(.)}\esp_{\prob_{Y_0\,|\,Y_{-k}^{-1}}}\left[ \rhotau \left(Y_0 - q(Y_{-k}^{-1})\right)\right].$$
Observe that $\{Z_k\}_{k =0}^{\infty}$ is a nonnegative supermartingale with respect to the family of sigma algebras $\{\sigma(Y_{-k}^{-1})\}_{k=1}^\infty$. In addition,
\begin{align*}
\sup_{k \geq 1}\esp[ Z_k^2] &= \sup_{k \geq 1}\esp\left[\left( \min_{q(.)} \esp_{\prob_{Y_0\,|\,Y_{-k}^{-1}}} \rho_\tau\left(Y_0 - q(Y_k^{-1})\right) \right)^2 \right] \\
& \leq \sup_{k \geq 1}\esp\left[\left( \esp_{\prob_{Y_0\,|\,Y_{-k}^{-1}}} \rho_\tau\left(Y_0\right)\right)^2 \right] \\
& \leq \sup_{k \geq 1} \esp \left[ \left(\rho_\tau(Y_0 )\right)^2\right] \\
& \quad \mbox{(by Jensen's inequality)}\\
& \leq \sup_{k \geq 1} \esp [ Y_0 ^2] < \infty.
\end{align*}
Therefore,
$$\mathbb E [Z_k] \underset{k \to \infty }\to \mathbb E [Z_\infty],$$
where
$$Z_\infty = \min_{q(.)}\esp_{\prob_{Y_0 \,|\, Y_{-\infty}^{-1}}}\left[ \rhotau \left(Y_0 - q(Y_{-\infty}^{-1})\right)\right].$$

Consequently,
$$\lim_{k \to \infty} \varepsilon_k^{\star} = L^{\star}.$$
We finish the proof by using Lemma \ref{lem:Ineq}. On the one hand, a.s.,
\begin{align*}
&\limsup_{n \rightarrow \infty} \inf_{k,\ell} \left( L_n\left(h_n^{(k,\ell)} -\frac{2\ln b_{k,\ell}}{n\eta_{n+1}} \right) \right)\\
&\quad\leq \inf_{k,\ell}\left( \limsup_{n \rightarrow \infty} L_n\left(h_n^{(k,\ell)} -\frac{2\ln b_{k,\ell}}{n\eta_{n+1}} \right)\right) \\
&\quad\leq \inf_{k,\ell} \left( \limsup_{n \rightarrow \infty} L_n\left( h_n^{(k,\ell)}\right) \right)\\
&\quad= \inf_{k,\ell} \varepsilon_{k,\ell}\\
&\quad\leq \lim_{k \to \infty} \lim_{\ell \to \infty} \varepsilon_{k,\ell}\\
&\quad\leq L^{\star}.
\end{align*}
Moreover,
\begin{align*}
&\frac{1}{2n} \sum_{t=1}^{n}\eta_{t} \sum_{k,\ell = 1}^{\infty}{p_{k,\ell,n}\left[\rhotau \left(Y_t - h_t^{(k,\ell)}(Y_1^{t-1})\right)\right]^2} \\
&\quad \leq\frac{1}{2n} \sum_{t=1}^{n}\eta_{t} \sum_{k= 1}^{\infty}\sum_{\ell=1}^{\infty}p_{k,\ell,n}\left[\rhotau \left(Y_t - T_{\min(t^\delta,\ell)}\left(\bar{h}_t^{(k,\ell)}(Y_1^{t-1})\right)\right)\right]^2 \\
&\quad \leq \frac{1}{2n} \sum_{t=1}^{n}{\eta_{t} \sum_{k= 1}^{\infty}\sum_{\ell=1}^{\infty}{p_{k,\ell,n}{\left|Y_t - T_{\min(t^\delta,\ell)}\left(\bar{h}_t^{(k,\ell)}(Y_1^{t-1})\right)\right|^2}}}.
\end{align*}
Thus
\begin{align*}
&\frac{1}{2n} \sum_{t=1}^{n}\eta_{t} \sum_{k,\ell = 1}^{\infty}p_{k,\ell,n}\left[\rhotau \left(Y_t - h_t^{(k,\ell)}(Y_1^{t-1})\right)\right]^2 \\
&\quad \leq\frac{1}{n} \sum_{t=1}^{n}\eta_{t} \sum_{k= 1}^{\infty} \left(\sum_{\ell=1}^{\infty}p_{k,\ell,n}|Y_t|^2 + \sum_{\ell=1}^{\infty} p_{k,\ell,n}\left[ T_{\min(t^\delta,\ell)}\left(\bar{h}_t^{(k,\ell)}(Y_1^{t-1})\right)\right]^2 \right)\\
&\quad \leq \frac{1}{n} \sum_{t=1}^{n}{\eta_{t} \sum_{k= 1}^{\infty}\left(\sum_{\ell=1}^{\infty}{p_{k,\ell,n}{|Y_t|^2 + \sum_{\ell=1}^{\infty} p_{k,\ell,n} t^{2\delta}}}\right)} \\
&\quad  \leq \frac{1}{n} \sum_{t=1}^{n}\eta_{t} \sum_{k,\ell =1}^{\infty}{p_{k,\ell,n}(Y_t^2 + t^{2\delta})}\\
&\quad  = \frac{1}{n}\sum_{t=1}^n\eta_t(t^{2\delta} + Y_t^2).
\end{align*}

Therefore, since $n^{2\delta}\eta_n \to 0$ as $n \to \infty$ and $\esp [Y_0^2]< \infty$,
$$\limsup_{n\to \infty}\frac{1}{2n} \sum_{t=1}^{n}\eta_{t} \sum_{k,\ell = 1}^{\infty}p_{k,\ell,n}\left[\rhotau \left(Y_t - h_t^{(k,\ell)}(Y_1^{t-1})\right)\right]^2 = 0 \quad\mbox{a.s.}$$
Putting all pieces together, we obtain, a.s.,
$$ \limsup_{n \to \infty}L_n(g) \leq L^{\star},$$
and this proves the result.
\begin{flushright}
$\square$
\end{flushright}

\subsection{Proof of Lemma \ref{tilt}}
To prove the first statement of the lemma, it will be enough to show that, for all $q\in \mathbb R,$%
\[
\mathbb{E}\left[ \rho _{\tau }\left( Y-q\right) \right] -\mathbb{E}\left[
\rho _{\tau }\left( Y-q_{\tau }\right) \right] \geq 0.
\]%
We separate the cases $q\geq q_{\tau }$ and $q<q_{\tau}$.
\begin{enumerate}
\item[$(i)$] If $q\geq q_{\tau }$, then
\begin{align*}
&\mathbb{E}\left[ \rho _{\tau }( Y-q )\right] -\mathbb{E}\left[
\rho _{\tau }( Y-q_{\tau }) \right]  \\
&\quad =\mathbb{E}\left[ ( Y-q) ( \tau -\mathbf 1_{[Y\leq q]})
-( Y-q_{\tau })( \tau -\mathbf 1_{[Y\leq q_{\tau }]}) \right]\\
&\quad =\mathbb{E}\left[( Y-q) \left ( \tau -( \mathbf 1_{[Y\leq q_{\tau}]}+ \mathbf 1_{[q_{\tau }<Y\leq q]}) \right) -( Y-q_{\tau }) (\tau -\mathbf 1_{[Y\leq q_{\tau }]}) \right]  \\
&\quad =\mathbb{E}\left[ ( q_{\tau }-q) ( \tau -\mathbf 1_{[Y\leq q_{\tau}]}) \right] -\mathbb{E}\left[ ( Y-q) \mathbf 1_{[q_{\tau }<Y\leq q]}\right].
\end{align*}%
We have
\begin{align*}
\mathbb{E}\left[ ( q_{\tau }-q) ( \tau -\mathbf 1_{[Y\leq q_{\tau}]}) \right]  &= ( q_{\tau }-q)\left ( \tau -\mathbb P [ Y\leq q_{\tau }] \right)\\
& = ( q_{\tau }-q) \left [ \tau -F_Y \left(F^{\leftarrow}_Y(\tau)\right)\right]\\
&\geq 0
\end{align*}%
and, clearly,
\begin{equation*}
-\mathbb{E}\left[ ( Y-q) \mathbf 1_{[q_{\tau }<Y\leq q]}\right] \geq 0.
\end{equation*}
This proves the desired statement.
\item[$(ii)$] If $q<q_{\tau }$, then
\begin{align*}
&\mathbb{E}\left[ \rho _{\tau }\left( Y-q\right) \right] -\mathbb{E}\left[
\rho _{\tau }( Y-q_{\tau }) \right]  \\
&\quad =\mathbb{E}\left[ (Y-q) ( \tau -\mathbf 1_{[Y\leq q]})
-( Y-q_{\tau }) ( \tau -\mathbf 1_{[Y\leq q_{\tau }]}) \right]
\\
&\quad =\mathbb{E}\left[ ( Y-q) ( \tau -\mathbf 1_{[Y\leq q]})
-( Y-q_{\tau }) \left( \tau -( \mathbf 1_{[Y\leq q]}+\mathbf 1_{[q<Y\leq
q_{\tau }]}) \right) \right]  \\
&\quad =\mathbb{E}\left[ ( q_{\tau }-q) ( \tau -\mathbf 1_{[Y\leq
q]}) \right] -\mathbb{E}\left[ ( Y-q_{\tau }) ( \tau
-\mathbf 1_{[q<Y\leq q_{\tau }]}) \right].
\end{align*}%
For $q<q_{\tau }$, $\prob\left[ Y\leq q\right]=F_Y(q) <\tau $. Consequently
$$\mathbb{E}\left[ ( q_{\tau }-q) ( \tau -\mathbf 1_{[Y\leq
q]}) \right] > 0.$$
Since
$$-\mathbb{E}\left[ ( Y-q_{\tau }) ( \tau
-\mathbf 1_{[q<Y\leq q_{\tau }]}) \right] \geq 0,$$
we are led to the desired result.
\end{enumerate}
Suppose now that $F_Y$ is increasing. To establish the second statement of the lemma, a quick inspection of the proof reveals that it is enough to prove that, for $q>q_{\tau}$,
$$\mathbb{E}\left[ ( Y-q) \mathbf 1_{[q_{\tau }<Y\leq q]}\right] < 0.$$
Take $q'\in (q_{\tau},q)$ and set $S=[q_{\tau}<Y\leq q']$. Clearly,
$$\mathbb P(S)=F_Y(q')-F_Y(q_{\tau})>0.$$
Therefore
$$\mathbb{E}\left[ ( Y-q) \mathbf 1_{[q_{\tau }<Y\leq q]}\right] \leq \mathbb{E}[ ( Y-q) \mathbf 1_{S}] < 0.$$
\begin{flushright}
$\square$
\end{flushright}

\paragraph{Acknowledgments.} The authors are greatly indebted to Adrien Saumard and Joann\`es Vermorel for their valuable comments and insightful suggestions on the first draft of the paper. They also thank two referees and the Associate Editor for their careful reading of the paper and thoughtful critical comments.

\bibliographystyle{IEEEtran}
\bibliography{bibliobiaupatra}

\begin{thebibliography}{10}
\providecommand{\url}[1]{#1}
\csname url@samestyle\endcsname
\providecommand{\newblock}{\relax}
\providecommand{\bibinfo}[2]{#2}
\providecommand{\BIBentrySTDinterwordspacing}{\spaceskip=0pt\relax}
\providecommand{\BIBentryALTinterwordstretchfactor}{4}
\providecommand{\BIBentryALTinterwordspacing}{\spaceskip=\fontdimen2\font plus
\BIBentryALTinterwordstretchfactor\fontdimen3\font minus
  \fontdimen4\font\relax}
\providecommand{\BIBforeignlanguage}[2]{{%
\expandafter\ifx\csname l@#1\endcsname\relax
\typeout{** WARNING: IEEEtran.bst: No hyphenation pattern has been}%
\typeout{** loaded for the language `#1'. Using the pattern for}%
\typeout{** the default language instead.}%
\else
\language=\csname l@#1\endcsname
\fi
#2}}
\providecommand{\BIBdecl}{\relax}
\BIBdecl

\bibitem{BRO1}
P.~J. Brockwell and R.~A. Davis, \emph{Time Series: Theory and Methods},
  2nd~ed.\hskip 1em plus 0.5em minus 0.4em\relax New York: Springer-Verlag,
  1991.

\bibitem{GYO1}
L.~Gy{\"o}rfi, W.~H{\"a}rdle, P.~Sarda, and P.~Vieu, \emph{Nonparametric Curve
  Estimation from Time Series}.\hskip 1em plus 0.5em minus 0.4em\relax Berlin:
  Springer-Verlag, 1989.

\bibitem{BOS1}
D.~Bosq, \emph{Nonparametric Statistics for Stochastic Processes: Estimation
  and Prediction}.\hskip 1em plus 0.5em minus 0.4em\relax New York:
  Springer-Verlag, 1996.

\bibitem{GAN1}
A.~Gannoun, J.~Saracco, and K.~Yu, ``Nonparametric prediction by conditional
  median and quantiles,'' \emph{J. Statist. Plann. Inference}, vol. 117, pp.
  207--223, 2003.

\bibitem{KOE3}
R.~Koenker and K.~F. Hallock, ``Quantile regression,'' \emph{J. Eco. Persp.},
  vol.~15, pp. 143--156, 2001.

\bibitem{DAR1}
D.~Duffie and J.~Pan, ``An overview of value at risk,'' \emph{J. Derivatives},
  vol.~4, pp. 7--49, 1997.

\bibitem{KOE1}
R.~Koenker, \emph{Quantile Regression}.\hskip 1em plus 0.5em minus 0.4em\relax
  Cambridge: Cambridge University Press, 2005.

\bibitem{CES1}
N.~Cesa-Bianchi and G.~Lugosi, \emph{Prediction, Learning, and Games}.\hskip
  1em plus 0.5em minus 0.4em\relax New York: Cambridge University Press, 2006.

\bibitem{KOE2}
R.~Koenker and G.~Bassett, Jr., ``Regression quantiles,'' \emph{Econometrica},
  vol.~46, pp. 33--50, 1978.

\bibitem{ALG2}
P.~Algoet, ``The strong law of large numbers for sequential decisions under
  uncertainty,'' \emph{IEEE Trans. Inform. Theory}, vol.~40, pp. 609--633,
  1994.

\bibitem{GYO3}
L.~Gy{\"o}rfi and G.~Lugosi, ``Strategies for sequential prediction of
  stationary time series,'' in \emph{Modeling Uncertainty}, ser. Internat. Ser.
  Oper. Res. Management Sci.\hskip 1em plus 0.5em minus 0.4em\relax Boston:
  Kluwer Acad. Publ., 2002, vol.~46, pp. 225--248.

\bibitem{NOB1}
A.~B. Nobel, ``On optimal sequential prediction for general processes,''
  \emph{IEEE Trans. Inform. Theory}, vol.~49, pp. 83--98, 2003.

\bibitem{GYO5}
L.~Gy\"{o}rfi and G.~Ottucs\'{a}k, ``Sequential prediction of unbounded
  sationary time series,'' \emph{IEEE Trans. Inform. Theory}, vol.~53, pp.
  1866--1872, 2007.

\bibitem{BIA1}
G.~Biau, K.~Bleakley, L.~Gy\"{o}rfi, and G.~Ottucs\'{a}k, ``Nonparametric
  sequential prediction of time series,'' \emph{J. Nonparametr. Stat.},
  vol.~22, pp. 297--317, 2010.

\bibitem{ALG1}
P.~Algoet, ``Universal schemes for prediction, gambling and portfolio
  selection,'' \emph{Ann. Probab.}, vol.~20, pp. 901--941, 1992.

\bibitem{GYO6}
L.~Gy\"{o}rfi and D.~Sch\"{a}fer, ``Nonparametric prediction,'' \emph{Advances
  in Learning Theory: Methods, Models and Applications}, pp. 341--356, 2003.

\bibitem{GYO4}
L.~Gy{\"o}rfi, G.~Lugosi, and F.~Udina, ``Nonparametric kernel-based sequential
  investment strategies,'' \emph{Math. Finance}, vol.~16, pp. 337--357, 2006.

\bibitem{GYO8}
L.~Gy{\"o}rfi, F.~Udina, and H.~Walk, ``Nonparametric nearest neighbor based
  empirical portfolio selection strategies,'' \emph{Statist. Decisions},
  vol.~26, pp. 145--157, 2008.

\bibitem{GYO2}
L.~Gy{\"o}rfi, M.~Kohler, A.~Krzy{\.z}ak, and H.~Walk, \emph{A
  Distribution-Free Theory of Nonparametric Regression}.\hskip 1em plus 0.5em
  minus 0.4em\relax New York: Springer-Verlag, 2002.

\bibitem{BUN1}
F.~Bunea and A.~Nobel, ``Sequential procedures for aggregating arbitrary
  estimators of a conditional mean,'' \emph{IEEE Trans. Inform. Theory},
  vol.~54, pp. 1725--1735, 2008.

\bibitem{GYO7}
\BIBentryALTinterwordspacing
L.~Gy\"orfi, F.~Udina, and H.~Walk, ``Experiments on universal portfolio
  selection using data from real markets,'' 2008, technical report. [Online].
  Available: \url{http://tukey.upf.es/papers/NNexp.pdf}
\BIBentrySTDinterwordspacing

\bibitem{CAR2}
J.~O. Street, R.~J. Carroll, and D.~Ruppert, ``A note on computing robust
  regression estimates via iteratively reweighted least squares,'' \emph{Amer.
  Statistician}, vol.~42, pp. 152--154, 1988.

\bibitem{TAK1}
I.~Takeuchi, Q.~V. Le, T.~D. Sears, and A.~J. Smola, ``Nonparametric quantile
  estimation,'' \emph{J. Mach. Learn. Res.}, vol.~7, pp. 1231--1264, 2006.

\bibitem{HAL1}
P.~Hall, L.~Peng, and Q.~Yao, ``Prediction and nonparametric estimation for
  time series with heavy tails,'' \emph{J. Time Ser. Anal.}, vol.~23, pp.
  313--331, 2002.

\bibitem{MAD1}
S.~G. Madrikakis, S.~C. Wheelwright, and R.~J. Hyndman, \emph{Forecasting,
  Methods and Applications}.\hskip 1em plus 0.5em minus 0.4em\relax New York:
  Wiley, 1998.

\bibitem{BRE1}
L.~Breiman, ``The individual ergodic theorem of information theory,''
  \emph{Ann. Math. Statist.}, vol.~28, pp. 809--811, 1957.

\bibitem{BIL1}
P.~Billingsley, \emph{Probability and Measure}, 3rd~ed.\hskip 1em plus 0.5em
  minus 0.4em\relax New York: John Wiley \& Sons, 1995.

\end{thebibliography}

\end{document}

\end{proof}

\end{document}